\documentclass{aa}
\usepackage{times,graphics}
\usepackage{lscape}

\def\teff{T$_{\rm eff}$}

\def\etal{{\em et~al}.  }
\def\km{kms$^{-1}$}
\def\ie{{\em i.e}.~}
\usepackage{epsfig} 
\begin{document}
   \title{Understanding B-type supergiants in the low metallicity environment of
   the SMC II.}
   
   \author{C. Trundle$^{1,2}$ \& D.J. Lennon$^{1}$} 

   \offprints{C.Trundle at ctrundle@ll.iac.es. \\
   Recently moved to Instituto de Astrofis\'ica de Canarias, C/ V\'ia L\'actea, E-38200 La Laguna
   Tenerife, Spain.}

   \authorrunning{C.Trundle \& D.J. Lennon}
   
   \titlerunning{SMC B-type Supergiants II}
    
   \institute{The Isaac Newton Group of Telescopes,
	      Apartado de Correos 321, E-38700,
	      Santa Cruz de La Palma, Canary Islands, Spain
		\and 
	      The Department of Pure and Applied Physics,
              The Queen's University of Belfast,
	      Belfast BT7 1NN, Northern Ireland	       
              }
    
   \date{}

   \abstract{Despite a resurgence of effort over the last decade in the area of
massive stars there is still ambiguity over their evolutionary path,
contamination of their surface abundances and the behaviour of their stellar
winds. Here 10 SMC B-type supergiants are analysed applying a unified model
atmosphere code {\sc fastwind} to intermediate resolution spectra from the ESO
Multi Mode Instrument ({\sc emmi}) on the NTT telescope. Combined with the 8
targets analysed in paper 1 (Trundle \etal 2004), this work provides
observational results on the properties of the winds and chemical compositions
of B-type supergiants in the SMC. This paper emphasizes and substantiates
the implications for stellar evolution from paper 1; that current theoretical
models need to produce larger degrees of surface nitrogen enhancements at lower
rotational velocities.  In addition a significant discrepancy between
theoretical and observed mass-loss rates is discussed which will have important
implications for the rotational velocities obtained from stellar evolution
calculations. Furthermore, an initial calibration of the wind-momentum
luminosity relationship for B-type supergiants in a low metallicity environment
(Z = 0.004) is presented.}

   \maketitle 


\section{Introduction}

The evolution of a galaxy is strongly influenced by the input of energy,
momentum and processed material from massive stars into the interstellar medium
(ISM). As possible progenitors of Type II supernovae and gamma ray bursts, the
nucleosynthesis processes and evolution of massive stars are important topics
in current astrophysical research. In spite of its importance, there is still a
large degree of ambiguity concerning the evolutionary path of these objects
beyond the main-sequence and the influence of rotation and mass-loss on the
composition of their stellar photospheres. Hence quantitative observational
studies concerning the chemical compositions and winds of massive stars are
vital tests on both the theory of galactic and stellar evolution. These
early-type stars are also important contributors to the spectra of 
star-forming galaxies and intermediate age (20-50 Myr) stellar populations (de
Mello et al. 2000). Furthermore in order to model the integrated light of
starburst regions and constrain the composition of high redshift galaxies,
accurate atmospheric and wind parameters of OB stars are required (Leitherer et
al. 1999; Pettini et al. 2000).

In the mid 1990's an intriguing implication arose from the theory of
radiatively driven winds and was introduced by Kudritzki, Lennon \& Puls
(1995). This work showed that the wind-momentum of a star, a product of its
mass-loss, terminal velocity and radius, is directly related to its luminosity
and can therefore provide an estimate of the stellar distance. The
intrinsically high luminosity of early-type supergiants coupled with
their relatively normal spectral behaviour makes them the best candidates for
using this Wind-momentum Luminosity Relationship (WLR). Recent work has shown
the dependence of the WLR on spectral type for Galactic OBA-type stars (Puls
\etal 1996; Kudritzki \etal 1999; Repolust, Puls \& Herrero 2004; Markova \etal
2004). Some studies have also endeavoured to understand the behaviour of the
WLR in the lower metallicity environments of the Magellanic Clouds (Puls \etal
1996; Kudritzki \& Puls 2000; Crowther \etal 2002; Trundle \etal 2004;  Evans
\etal 2004a). Most of these studies focused on relatively
small samples of O-type stars with some indication of the WLR having different
behaviour at lower metallicity. Recent work by Trundle \etal (2004; hereafter
paper 1) investigated a sample of B-type supergiants in the SMC. Although the
sample only included 8 targets, it provided observational evidence to test line
driven wind theory in this spectral region and at a low metallicity. Indeed
this work highlighted a significant discrepancy between the observed mass-loss
rates and the theoretical predictions by Vink \etal (2001) at SMC
metallicities.

\begin{table*}
\caption[SMC {\sc emmi} Observational Details.]{\label{smc2obs} Observational 
details. Identification numbers are from Azzopardi \& Vigneau (1982; AV\#) and
Sanduleak (1968; Sk\#). Spectral types adopted from Lennon (1997), except for
AV78 which has been reclassified in this work from the blue spectra.  Absolute
Magnitudes (M$_{v}$) are calculated using V and (B-V) magnitudes from Garmany,
Conti \& Massey (1987;$^{1}$) and Massey (2002;$^{2}$) and (B-V)$_{0}$ values
from Fitzpatrick \& Garmany (1990). The adopted distance modulus is 18.9
(Harries, Hilditch, \& Howarth 2003). The average S/N ratios for the blue and
red arms are presented. The $v\sin i$ values  represent the width of the
spectral features and are simply upper limits on the projected rotational
velocities (see Paper 1).
}
\begin{flushleft}
\centering
\begin{tabular}{lllccccccc} \hline \hline
STAR & Alias & Spectral & V &  B - V & M$_{v}$ & \multicolumn{2}{c}{S/N} & v$_{\rm lsr}$ & v$_{\sin i}$  \\
     &       & Type     &   &        &         &    B  & R &        (kms$^{-1}$) & (kms$^{-1}$)\\ 
\hline   
\\            
AV420 & Sk131 & B0.5 {\sc I}a      & 13.09$^{2}$ &-0.17 & -5.90 &  84 &  49 & 188 $\pm$ 12 & 80
\\
AV242 & Sk85  & B1 {\sc I}a        & 12.11$^{1}$ &-0.13 & -6.91 & 125 &  47 & 188 $\pm$ 25 & 90
\\
AV264 & Sk94  & B1 {\sc I}a        & 12.36$^{1}$ &-0.15 & -6.60 & 102 &  74 & 142 $\pm$ 14 & 85
\\ 
AV78  & Sk40  & B1 {\sc I}a$^{+}$  & 11.05$^{1}$ &-0.03 & -8.25 & 161 &  80 & 176 $\pm$ 12 & 85
\\
AV96  & Sk46  & B1.5 {\sc I}a      & 12.59$^{2}$ &-0.10 & -6.50 & 146 &  69 & 161 $\pm$ 10 & 90
\\   
AV373 & Sk119 & B2   {\sc I}a      & 12.17$^{1}$ &-0.09 & -6.92 &  80 & 105 & 191 $\pm$ 26 & 80
\\   
AV10  & Sk7   & B2.5 {\sc I}a      & 12.58$^{2}$ &-0.02 & -6.69 &  80 & 110 & 225 $\pm$ 21 & 85
\\   
AV56  & Sk31  & B2.5 {\sc I}a      & 11.15$^{1}$ &-0.00 & -8.18 & 120 & 100 & 209 $\pm$ 11 & 80
\\   
AV443 & Sk137 & B2.5 {\sc I}a      & 10.97$^{1}$ &-0.06 & -8.18 & 165 & 160 & 191 $\pm$ 22 & 73
\\   
AV151 & Sk57  & B2.5 {\sc I}a      & 12.26$^{2}$ &-0.02 & -7.01 & 122 &  80 & 187 $\pm$ 15 & 62
\\   
\hline             
\end{tabular}
\end{flushleft}
\end{table*} 

In addition to wind analyses, Paper 1 investigated the chemical composition of
the SMC sample. This provided quantitative CNO abundances of B-type
supergiants; a key test of the mixing and mass-loss processes incorporated in
stellar evolution models for early-type stars. The results from Paper 1 are
consistent with the significant enhancements and dispersions of nitrogen
abundances,  previously observed in B-type stars (Gies \& Lambert 1992; Lennon
et al. 1991, 1996, 1997, 2003; Fitzpatrick \& Bohannan 1993; McErlean, Lennon,
\& Dufton 1999; Dufton et al. 2000). The existence of a stellar wind can reduce
the angular momentum of a star, which in turn causes a decrease in the
rotational velocity. This link between the mass-loss of a star and its
rotational velocity affects the stellar lifetimes and photospheric composition
predicted by stellar evolution codes. Thus mass-loss rates, rotational
velocities and abundances are important observational constraints for such
codes. The recent inclusion of rotationally induced mixing in stellar
evolution codes such as those by Maeder \& Meynet (2001), were shown in paper 1
to reproduce the large dispersion of nitrogen abundances observed in B-type
supergiants, implying that rotation may play a significant role in the
evolution of massive stars.  

The dataset of Lennon (1997) offers the opportunity of analysing an additional
ten B-type supergiants which supplements the results from paper 1 and will
provide a clearer insight into the behaviour of these luminous objects. The
objective of this work is to provide observational constraints on the mass-loss
rates of B-type supergiants and to calibrate the WLR at the metallicity of the
SMC.

\section{Observational Data}
\label{obs2}

The SMC optical dataset considered for this work is a subset of that presented
by Lennon (1997), in a study to delineate the spectral classifications of
B-type supergiants in low metallicity environments. The medium resolution
spectra (R $\sim$ 20000) were obtained remotely on the NTT telescope with the
ESO Multi Mode Instrument ({\sc emmi}). Three spectral regions were observed
covering the wavelengths 3925-4375 \AA, 6190-6830 \AA\ and 4300-4750 \AA. 
Signal-to-noise (S/N) ratios greater than 70 were obtained along with
dispersions of 0.45 and 0.32 \AA\ per pixel in the blue and red arms,
respectively  (see Table~\ref{smc2obs}).  Further details of the observations
and data reduction procedures can be found in Lennon (1997).

The original data selected by Lennon (1997) are a subset of B0 - B9 stars from
the sample of Garmany \etal (1987), who carried out a spectral classification
of SMC OB-type stars. For the purpose of this paper ten stars from this dataset
have been selected spanning the spectral range B0 - B2.5. An additional
constraint was that each object have medium to strong stellar winds, which was
apparent from the profile of the H$_{\alpha}$ line. In Table~\ref{smc2obs} the
targets are listed with the spectral types assigned by Lennon (1997) from
applying his revised classifications for B-type supergiants in the SMC. AV78
has been reclassified here as a B1 Ia$^{+}$ star due to the presence of a weak
Si {\sc iv} line in its spectrum at 4116 \AA. This star was previously classed
as a B1.5 Ia$^{+}$ star by Lennon (1997). 

The estimated projected rotational velocities ($v\sin i$) were determined using
the techniques described in paper 1 and are presented in Table~\ref{smc2obs}.
It is important to note that the $v\sin i$ values only represent the width of
the spectral features and are simply upper limits for the projected rotational
velocities (\ie no attempt has been made to deconvolve the contribution of
stellar rotation and macroturbulent motions from the line broadening).
Equivalent widths were measured by fitting a Gaussian profile using a
non-linear least squares technique in {\sc dipso}  (Howarth et al. 2003); a
spectrum analysis package from {\sc starlink}. Table~\ref{smc2obs} also lists
the absolute magnitudes calculated using V and (B-V) magnitudes from Garmany,
Conti \& Massey  (1987) and Massey (2002) and (B-V)$_{\rm 0}$
values from Fitzpatrick \& Garmany (1990). The adopted distance modulus of the
SMC, applied to determine the absolute magnitude (M$_{v}$), is 18.9 (Harries,
Hilditch, \& Howarth 2003). 

Radial velocities of the target stars are corrected to the local standard of
rest, giving a range in velocities of 142 - 225 kms$^{-1}$ (note similar range
observed for SMC stars in Paper 1). From a H {\sc i} 21 cm emission survey by
McGee \& Newton (1981), four gas complexes were identified in the SMC at
heliocentric velocities of 114, 134, 167 \& 192 kms$^{-1}$. In addition Welty
\etal (1997) found SMC multiple absorption components of interstellar gas with
velocities between 85 $\leq$ and 210 kms$^{-1}$.  This implies that the radial
velocities of our target stars are consistent with those of field
stars in the SMC. 

\section{Stellar parameters} 
\label{smc2param}

\begin{figure*}[ht]
\center{
\resizebox{134mm}{235mm}{\includegraphics{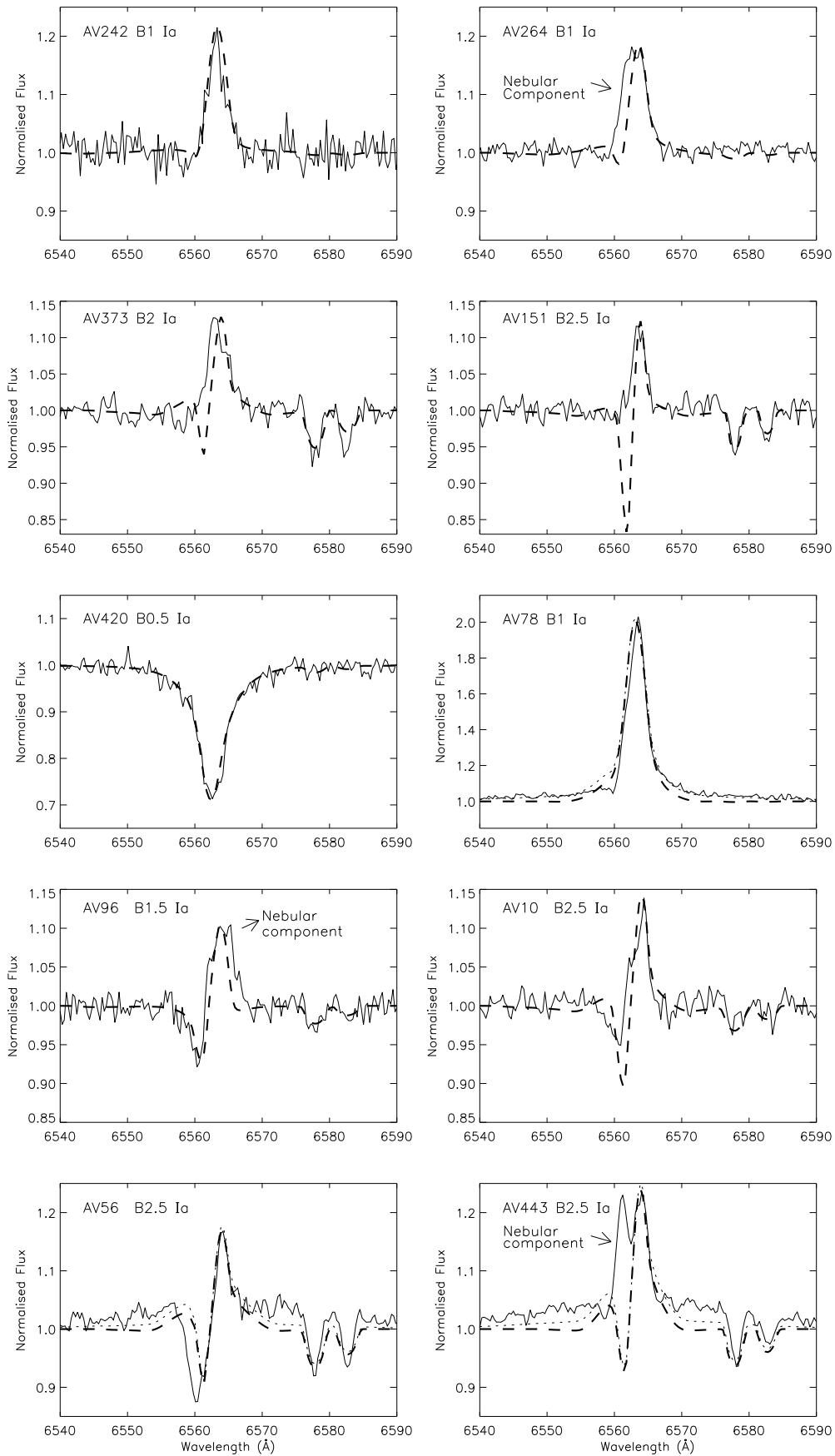}}
\caption{\label{ha2ab}  H$_{\alpha}$ profiles for the ten SMC B-type
supergiants. The solid-dashed lines ({\bf- - -}) are the best fits from the
{\sc fastwind} models and the parameters are given in Table~\ref{smc2par3}. In
the case of AV78, AV56 \& AV443, models with the best fit parameters but
with incoherent electron scattering included in the formal solution are also
shown (dotted line; $\cdot\cdot\cdot$).}} \end{figure*}

The model atmospheres and procedures used in this analysis are identical to
those described in Paper 1 and will only briefly be repeated here. The `unified
model atmosphere' code applied is called {\sc fastwind}, a line-blanketed,
spherically symmetric, non-LTE code (Santolaya-Rey, Puls, \& Herrero (1997);
Herrero, Puls, \& Najarro 2002; Repolust, Puls, \& Herrero 2004). The
temperature structure in {\sc fastwind} was not calculated explicitly, but
parameterised by applying the non-LTE Hopf function described by Santolaya-Rey,
Puls, \& Herrero (1997; see equations 4.1 - 4.3) to {\sc tlusty} models
calculated at Queen's University Belfast (Ryans, R.S.I.; private
communication). 

The stellar parameters were calculated using precisely the same methods as
described in Paper 1. In short, the parameter space in which each star resides
was estimated using a coarse grid and this was followed by an iterative process
to determine each parameter accurately in a finer but smaller grid. The
resulting photospheric and wind parameters for each target are presented in
Table~\ref{smc2par3}. 

For the B0.5 - B2 stars, effective temperatures were determined from the Si
{\sc iii}/Si {\sc iv} ionisation balance with typical random errors of 1.5 kK.
In the B2 - B2.5 stars only the Si {\sc iii} lines were present, as a result
profile fitting to these lines was required to determine the temperature. For
these objects, the lack of a Si {\sc iv} 4116 \AA\ line and the very weak
features for the Si {\sc ii} 4128, 4131 \AA\ profiles acted as upper and lower
limits for the temperature estimates. The only exception to this is AV151 which
exhibited relatively strong Si {\sc ii} features allowing the Si {\sc ii}/Si
{\sc iii} ionisation balance to be determined. The uncertainty for AV151 is 1.5
kK, and is slightly larger for the other B2-B2.5 stars (2.0 kK). This
uncertainty in effective temperature also affects the determination of the
logarithmic surface gravity, typically an error of 0.5 kK represents a
systematic error of 0.05 dex in $\log$ g. In addition to this there is a random
error of $\pm$ 0.05 for the B2-B2.5 stars and up to 0.1 dex for the B0-B1.5
stars (See Table~\ref{smc2par3}).

Microturbulent velocities, $\xi_{\rm Si}$, adopted for the analysis were
derived from the Si {\sc iii} triplet at 4560 \AA. The microturbulence was also
derived from the O {\sc ii} multiplets, in the stars with spectral type earlier
than B2. As discussed in paper 1 these values tend to be $\sim$ 10 \km~ higher
than estimated from the silicon lines. In the case of the stars cooler than B2,
no O {\sc ii} lines are present in the spectra as a result of the low
temperatures and hence low ionisation of this species. For AV56 the spread in
equivalent widths amongst the Si {\sc iii} multiplet was only $\sim$ 50 m\AA\
and this was not sufficient to provide an accurate microturbulent velocity
(typically this spread is $>$ 100 m\AA\ in the rest of the sample). In addition
AV56 is too cool for it to have a well developed O {\sc ii} spectrum,
preventing an estimation of $\xi_{\rm O}$ and therefore a value of 10 \km~ was
adopted in the analysis for this star.  

Three targets, AV242, AV264 \& AV78, were included in the UV analysis of  Evans
\etal  (2004 - hereafter EVANS04) (described in Paper 1), using archive
IUE/HIRES data for AV78 and HST/GHRS spectra for the former two stars. These
objects, therefore have predetermined terminal velocities measured with the SEI
(Sobolev Exact Integration) method from the UV resonance lines. The remaining
seven stars have no available UV data from which terminal velocities can be
measured and hence estimates were made using the alternative procedure
described in Paper 1 \& EVANS04. This involves calculating the escape velocity
and applying the ratios of v$_{\infty}$/v$_{\rm esc}$ determined by Kudritzki
\& Puls (2000) (viz. for \teff $\geq$ 21 kK this ratio is equal to 2.65 and for
\teff $<$ 21 kK it is 1.4).

H$_{\alpha}$ profiles provide a means to determine the properties of the
stellar wind; with a knowledge of the terminal velocity, the mass-loss rate
(\.M)  and beta parameter ($\beta$) can be determined simultaneously  from the
profile shape. \.M controls the overall wind emission in the profile, whilst
$\beta$ causes a variation in the strength and FWHM of the emission peak. As
such it is difficult to constrain the beta parameter in stars with weak winds,
where wind emission may only cause a small amount of filling in of the
absorption core (viz. AV420; see Fig~\ref{ha2ab}). In fact a variation of
$\beta$ from 0.8 to 2 for AV420 with subsequent changes in \.M from  0.49
to 0.14.10$^{-6}$ M$_{\odot}$ yr$^{-1}$ will reproduce the observed
H$_{\alpha}$ profile. In this analysis we adopt a $\beta$ parameter of 1 for
this weak wind star with the knowledge that this will lead to uncertain
mass-loss rates (see also AV216 and AV104 in Paper 1). 

For the P-Cygni and emission profiles, $\beta$ can be determined rather more
accurately on the order of $\pm$ 0.5. This is considerably higher than the error
found from the {\sc uves} data and is attributed to the lower resolution and in
some cases lower S/N of the current dataset. In addition, nebular contributions in
AV96 \& AV264 prevent accurate estimates of the emission peak FWHM and a $\beta$
in the range 2 - 3 can generally fit these profiles with values above 3 having
little impact on the profile shape.  Since \.M and $\beta$ are derived
simultaneously from the fitting of H$_{\alpha}$ the error in the mass-loss rate is
dominated by this uncertainty in $\beta$. The difference in \.M derived  with a
$\beta$ of 2 and of 3 is $\sim$ 25 \%. In the case of AV264 a value of 3 was
deemed more appropriate from a fit to the redward wing of the emission profile and
the height of the emission peak. $\beta$ parameters for AV373 \& AV151 are as
uncertain as that for AV96 due to the failures in the code to produce complete
emission profiles at low temperatures (see below) and which contribute to the
error in the mass-loss rate derived. Uncertainties in the mass-loss rates
therefore range from 15\% to 25\% and are presented in Table~\ref{smc2par3}.

The final model fits to the observed H$_{\alpha}$ profiles are presented in
Figs.~\ref{ha2ab}. Three of the objects, AV78, AV443 \& AV56 display strong
broadening in their profile wings (similar to AV362 \& AV22 in paper 1). These
wings, which are formed in the upper layers of the photosphere, are
considerably affected by incoherent electron scattering. The introduction of
electron scattering into the formal solution of the {\sc fastwind} models
provides an excellent profile fit to AV78 (B1Ia$^{+}$; see Fig.~\ref{ha2ab}).

\begin{landscape}
\begin{center}
\begin{table}
\caption[Atmospheric and wind parameters of SMC supergiants]
{\label{smc2par3}  Derived atmospheric and wind parameters for SMC
B-type supergiants. The microturbulence from both the Si {\sc iii} \& O
{\sc ii} lines are presented, however the silicon mircoturbulence was
adopted in the analysis. A microturbulence of 10 kms$^{-1}$ was adopted 
for  AV56 due to lack of a direct method of measurement. Escape
velocities are calculated using the procedure outlined in paper 1 \&
EVANS04. Terminal velocities, (v$_{\infty}$), for AV242, AV264 \& AV78
are derived from the SEI method (1). For the rest of 
the sample,
v$_{\infty}$ is calculated by adopting a $\frac{v_{\infty}}{v_{\rm
esc}}$ ratio of 2.65 for stars with T$_{\rm eff}$ $\geq$ 21.0 kK and
1.40 for T$_{\rm eff}$ $<$ 21.0 kK (2). The errors 
quoted are discussed in
Section~\ref{smc2param}.}  
\begin{flushleft}
\centering
\begin{tabular}{lccccccccccccc} \hline \hline
Star &T$_{\rm eff}$ & $\log$ g & R$_{\star}$&M$_{\rm spec}$ &M$_{\rm evol}$ &$\log(\frac{L_{\star}}{L_{\odot}})$ & $\xi$$_{\rm Si}$ & $\xi$$_{\rm O}$ &  \.{M}& v$_{\infty}$  & v$_{\rm esc}$ & $\beta$ & $\log$(D$_{\rm MOM}$) \\	  
       &  (kK)         &   (cgs) & (R$_{\odot}$)    &(M$_{\odot}$)&(M$_{\odot}$)&			       & \multicolumn{2}{c}{ (kms$^{-1}$)}    &  (10$^{-6}$ M$_{\odot}$ yr$^{-1})$ &(kms$^{-1}$) &	    (kms$^{-1}$) &  &(cgs)\\					    
\hline   
\\            
AV420 & 27.0 $\pm$ 1.5 & 3.05 $\pm$ 0.15 & 21.7 & 19 & 26   & 5.35 & 13 & 25  & 0.34  $\pm$ 0.15  &  1310 $\pm$ 260 $^{2}$& 493 $^{2}$& 1.0	& 28.11 $\pm$ 0.40
\\
AV242 & 25.0 $\pm$ 1.5 & 2.85 $\pm$ 0.15 & 36.6 & 35 & 39.5 & 5.67 & 13 & 20  & 0.84 $\pm$ 0.13 &  950 $\pm$ 100 $^{1}$ & 359 $^{1}$& 2.0~ & 28.48 $\pm$ 0.15	  
\\
AV264 & 22.5 $\pm$ 1.5 & 2.55 $\pm$ 0.15 & 34.8 & 16 & 29   & 5.44 & 13 & 19  & 0.29  $\pm$ 0.06  &  600 $\pm$ 100 $^{1}$ & 313 $^{1}$& 2.5~ & 27.81 $\pm$ 0.20
\\
AV78  & 21.5 $\pm$ 1.5 & 2.40 $\pm$ 0.15 & 79.0 & 57 & 53   & 5.92 & 12 & 20  & 2.29  $\pm$ 0.34  &  450 $\pm$ ~50 $^{1}$ & 423 $^{1}$& 3.0~ & 28.76 $\pm$ 0.15
\\
AV96  & 22.0 $\pm$ 1.5 & 2.55 $\pm$ 0.15 & 34.0 & 15 & 27   & 5.39 & 11 & 21  & 0.24 $\pm$ 0.06  &  850 $\pm$ 170 $^{2}$ & 320 $^{2}$& 3.0~ & 27.87 $\pm$ 0.25   
\\   
AV373 & 19.0 $\pm$ 2.0 & 2.30 $\pm$ 0.20 & 46.8 & 16 & 28   & 5.42 & 11 & 22  & 0.16 $\pm$ 0.04  &  390 $\pm$ ~80 $^{2}$ & 282 $^{2}$& 3.0~ & 27.42 $\pm$ 0.25    
\\   
AV10  & 17.0 $\pm$ 2.0 & 2.20 $\pm$ 0.20 & 46.7 & 13 & 22   & 5.21 & 14 &     & 0.15  $\pm$ 0.03  &  380 $\pm$ ~75 $^{2}$ & 273 $^{2}$& 3.0~ & 27.39 $\pm$ 0.20  
\\   
AV56  & 16.5 $\pm$ 2.0 & 2.05 $\pm$ 0.20 & 96.1 & 38 & 50   & 5.88 & 10 &     & 0.51  $\pm$ 0.08  &  420 $\pm$ ~85 $^{2}$ & 302 $^{2}$& 2.0~ & 28.12 $\pm$ 0.15
\\   
AV443 & 16.5 $\pm$ 2.0 & 1.95 $\pm$ 0.20 & 96.5 & 30 & 42   & 5.79 & 11 &     & 0.45 $\pm$ 0.09 &  340 $\pm$ ~70 $^{2}$ & 246 $^{2}$& 2.0~ & 27.97 $\pm$ 0.20  
\\   
AV151 & 16.0 $\pm$ 1.5 & 2.10 $\pm$ 0.15 & 57.1 & 15 & 24   & 5.28 & 15 &     & 0.16  $\pm$ 0.04  &  370 $\pm$ ~85 $^{2}$ & 263 $^{2}$& 3.0~ & 27.45 $\pm$ 0.25
\\   
\hline 
\end{tabular}
\end{flushleft}
\end{table}   
\end{center}
\end{landscape}

\begin{table*}
\caption[Non-LTE absolute abundances of SMC 
supergiants]{\label{smc2abund}
Derived non-LTE absolute abundances for SMC B-type supergiants.
Abundances are given as $\log$ [N(X)/N(H)] + 12, where X represents the
element under consideration. The errors represent the standard
deviation of the mean and account for systematic errors. The 
numbers in the parenthesis are the number of lines included in the 
analysis. Mean element abundances are presented in the final line 
of the table. }
  \begin{flushleft}
\centering
\begin{tabular}{lccccccc} \hline \hline
Star &  C {\sc ii} & N {\sc ii}       & O {\sc ii}       & Mg {\sc 
ii}      & Si {\sc ii}& Si {\sc iii}     & Si {\sc iv}
\\
\hline 
AV420 &            $<$7.08 (1)& 7.44 $\pm$ 0.14 (2)& 8.06 $\pm$ 0.27 (10)& 6.84 $\pm$ 0.24 (1)&   & 6.62 $\pm$ 0.12 (3)& 6.59 $\pm$ 0.46 (1)
\\ 
AV242 &            $<$6.88 (1)& 7.40 $\pm$ 0.19 (1)& 8.24 $\pm$ 0.27 (8) & 6.84 $\pm$ 0.38 (1)&   & 6.87 $\pm$ 0.10 (3)& 6.91 $\pm$ 0.30 (1)
\\
AV264 &            $<$6.61 (1)& 7.86 $\pm$ 0.13 (3)& 8.01 $\pm$ 0.25 (13)& 6.91 $\pm$ 0.14 (1)&   & 6.65 $\pm$ 0.14 (3)& 6.64 $\pm$ 0.28 (1)	       
\\
AV78  &            $<$6.90 (1)& 8.30 $\pm$ 0.24 (7)& 7.86 $\pm$ 0.16 (12)& 6.95 $\pm$ 0.16 (1)&   & 7.22 $\pm$ 0.11 (3)& 7.26 $\pm$ 0.45 (1)
\\
AV96  &    6.88 $\pm$ 0.22 (3)& 7.67 $\pm$ 0.23 (8)& 8.09 $\pm$ 0.22 (13)& 6.97 $\pm$ 0.25 (1)&   & 6.85 $\pm$ 0.21 (3)&       $<$ 6.79 :(1)  
\\   
AV373 &    6.89 $\pm$ 0.19 (3)& 7.46 $\pm$ 0.25 (2)& 8.37 $\pm$ 0.39 (8)& 6.94 $\pm$ 0.30 (1)& $<$6.79 (2)& 6.74 $\pm$ 0.15 (3)&
\\   
AV10  &    6.59 $\pm$ 0.14 (3)& 7.66 $\pm$ 0.26 (3)& 8.13 $\pm$ 0.31  (3)& 6.81 $\pm$ 0.12 (1)& $<$6.76 (2)& 6.69 $\pm$ 0.16 (3)&
\\    
AV56  &    7.19 $\pm$ 0.53 (3)& 8.27 $\pm$ 0.32 (2)&    	         & 6.68 $\pm$ 0.09 (1)& $<$6.60 (2) & 6.64 $\pm$ 0.22 (3)& 
\\   
AV443 &    6.99 $\pm$ 0.36 (3)& 7.96 $\pm$ 0.22 (2)&      	         & 6.72 $\pm$ 0.08 (1)& $<$6.55 (2) & 6.63 $\pm$ 0.18 (3)&
\\   
AV151 &    6.85 $\pm$ 0.12 (3)& 7.55 $\pm$ 0.19 (2)&    	         & 6.70 $\pm$ 0.08 (1)&    6.54 $\pm$ 0.11 (2)& 6.61 $\pm$ 0.18 (3)&
\\   
\hline
Mean &  6.89 $\pm$ 0.18&  7.76 $\pm$ 0.28 & 8.11 $\pm$ 0.16 & 6.84 $\pm$ 0.11 & 6.65 $\pm$ 0.12 & 6.75 $\pm$ 0.18 & 6.84 $\pm$ 0.27 
\\
\hline
\end{tabular}
\end{flushleft} 
\end{table*}

\noindent However it doesn't completely account for the strong wings observed
in the two cooler B2.5Ia type stars; this was also the case for AV362 (B3Ia) \&
AV22 (B5Ia). Also note the models failure to reproduce the observed profiles
for spectral types later than B2 Ia (typically for \teff~ below 19kK). The
models below these temperatures exhibit strong absorption features not present
in the observed spectra. In these models, the Lyman continuum is optically
thick causing the transition from the 1$^{\rm st}$ to 2$^{\rm nd}$ energy
levels to be in detailed balance and the second energy level then simulates a
ground state. The H$_{\alpha}$ line (\ie transition from the 2$^{\rm nd}$ to
3$^{\rm rd}$ energy level) will subsequently result in a P-Cygni profile as it
behaves like a quasi-resonance line. This may, in part, be due to the current
temperature structures used in the {\sc fastwind} models and could possibly be
improved with the introduction of a temperature correction scheme. A new
version of {\sc fastwind} has recently been released which amongst other
advances incorporates such a scheme (Puls \etal 2005). It is based on an
alternative method introduced by Kub\'at, Puls \& Pauldrach (1999) to solve the
radiative equilibrium equation explicitly. With this technique, the temperature
in the upper photosphere and wind is calculated by ensuring that a thermal
balance of electrons is obtained.

One noticeable factor from Table 2 is the large values of the
$\beta$-parameter determined for these B-type objects. These differ from those
observed and predicted for O-type stars which have much faster winds with
$\beta$ being close to the order of unity. Indeed large $\beta$ values have
been observed in other optically based spectroscopic studies of B-type
supergiants in the Galaxy and SMC (Kudritzki \etal 1999; Evans \etal 2004a). On
the otherhand applying the SEI-method to UV P-Cygni lines of the SMC targets
presented here and in Paper 1 and of the Galactic targets analysed in Kudritzki
\etal suggest lower values of $\beta$ of 1-1.5 (EVANS04 and Haser 1995,
respectively). In both these UV analyses it was suggested that whilst small
changes of $\beta$ can give aesthetically more pleasing fits to the UV
resonance lines, they were relatively insensitive to the parameter. In fact,
Haser (1995) found that the SEI method of the B-type stars produce identical
profiles for $\beta$ values in the range 1-4. It is likely that the relatively
slow winds of these objects result in the P-Cygni lines under sampling the
velocity of the wind over the small range in depth for which the formation of
these lines cover, in comparison to the H$_{\alpha}$ profile. Hence the $\beta$
parameter is ill defined by the UV lines in these slower wind objects.

\section{Chemical Composition}

The procedures for the abundance analysis are identical to that discussed in
Paper 1, they apply the same atmospheric models and atomic data and as such one
would expect the abundances to follow the same patterns. Indeed, no surprises
are hidden in the chemical composition of these 10 B-type supergiants.
Consequently, a few interesting points will be introduced here with further
discussions of the implications in the following section. 

From a grid of equivalent widths generated by {\sc fastwind} models at various
microturbulent velocities and abundances, the absolute abundances of C, N, O,
Mg \& Si for the SMC stars were derived and are presented in
Table~\ref{smc2abund}. The dominating uncertainties in these abundances are the
adopted microturbulence, $\xi_{Si}$, equivalent widths and atomic data. In
addition the lower resolution of the {\sc emmi} data causes a slight {\it
relative} increase in the uncertainty of the equivalent width measurements,
particularly for closely spaced, weak lines which may have been resolved in the
{\sc uves} data but are slightly blended in the {\sc emmi} data (viz. Si {\sc
iv} 4116 \AA\ \& He {\sc i} 4121 \AA).

One difference in this work and that carried out for the {\sc uves} data are
the number of nitrogen and oxygen lines considered in the abundance analysis.
Experience with the B-type supergiant spectra identified some additional,
reliable lines to be included in the analysis (viz. O {\sc ii} 4069, 4349 \&
4676 and N {\sc ii} 4241 \AA). In general the weakest lines in the {\sc uves}
abundance analysis were not measurable in the {\sc emmi} data and were omitted
from the analysis.    

Following the procedures laid out in Paper 1, the carbon abundance
was corrected for NLTE effects due to the problematic line 4267 \AA.
Table~\ref{smc2abund} states a mean carbon abundance of 6.89 dex, by applying the
appropriate correction of +0.34 dex a mean C abundance of 7.23 dex is estimated
for this sample.  

It is worth noting, that preliminary abundance estimates for 64 SMC B-type
supergiants were made by Dufton \etal (2000) in a study of line strengths (this
work used the complete atlas of Lennon 1997). The Mg and Si results are in
close agreement with those found here, whilst the C, N and O abundances,
although lower differ by no more than 0.21 dex (see Table~\ref{smc2comp}). 
Dufton \etal found a range of nitrogen abundances from 7.09 to 7.69 dex to fit
the data, whilst the maximum value is smaller than that derived from the more
detailed analysis in this work there is some consistency between the spread in
results from the two analyses. This work shows that with a large sample, even
basic analysis techniques can arrive at reasonable estimates for the
chemical compositions of a stellar population.

\section{Discussion}
\label{smcchem2}

\begin{table*}
\caption[Mean non-LTE abundances of complete SMC supergiant 
sample.]{\label{smc2comp} Comparison of mean abundances for the complete
SMC B-type supergiant sample (including results from
paper 1) with previous studies of abundances in the
SMC. Included are the mean abundances of A-type
supergiants from Venn
\etal (1999, 2003; Si corrected for non-LTE effects.), SMC
main-sequence B-star AV304 (Rolleston \etal 2003; corrected for
non-LTE effects), NGC330 cluster main-sequence B-stars
(Lennon, Dufton \& Crowley 2003; corrected for non-LTE
effects) and SMC H {\sc ii}
regions from Kurt \etal (1999). The numbers in italics represent the abundances
predicted from EW measurements in SMC B-type supergiants by
Dufton \etal (2000; {\it D00}).
}  
\begin{flushleft}
\centering
\begin{tabular}{lllcccc} \hline \hline
        & \multicolumn{3}{c}{Supergiants}  & \multicolumn{2}{c}{B stars}& H{\sc ii} 
\\
 & This work & {\it D00} & A-type &  NGC330 & AV304 & regions
\\
\hline
\\
C   & 7.27 $\pm$ 0.14 & {\it 7.10} &  & 7.26 $\pm$ 0.15 & 7.41 $\pm$ 0.18 & 7.53 $\pm$ 0.06
\\
N   & 7.71 $\pm$ 0.32 & {\it 7.49} & 7.52 $\pm$ 0.10 & 7.51 $\pm$ 0.18 & 6.55 $\pm$ 0.01 & 6.59 $\pm$ 0.08
\\
O   & 8.13 $\pm$ 0.13 & {\it 7.95} & 8.14 $\pm$ 0.06 & 7.98 $\pm$ 0.13 & 8.16 $\pm$ 0.33 & 8.05 $\pm$ 0.05
\\
Mg  & 6.81 $\pm$ 0.14 & {\it 6.78} & 6.83 $\pm$ 0.08 & 6.59 $\pm$ 0.14 & 6.73            & 
\\
Si  & 6.75 $\pm$ 0.18 & {\it 6.78} & 6.92 $\pm$ 0.15 & 6.58 $\pm$ 0.32 & 6.74 $\pm$ 0.03 & 6.70 $\pm$ 0.20
\\
\hline 
\\
\end{tabular}
\end{flushleft}
\end{table*}

\begin{figure}
\center{
\resizebox{\hsize}{!}{\includegraphics{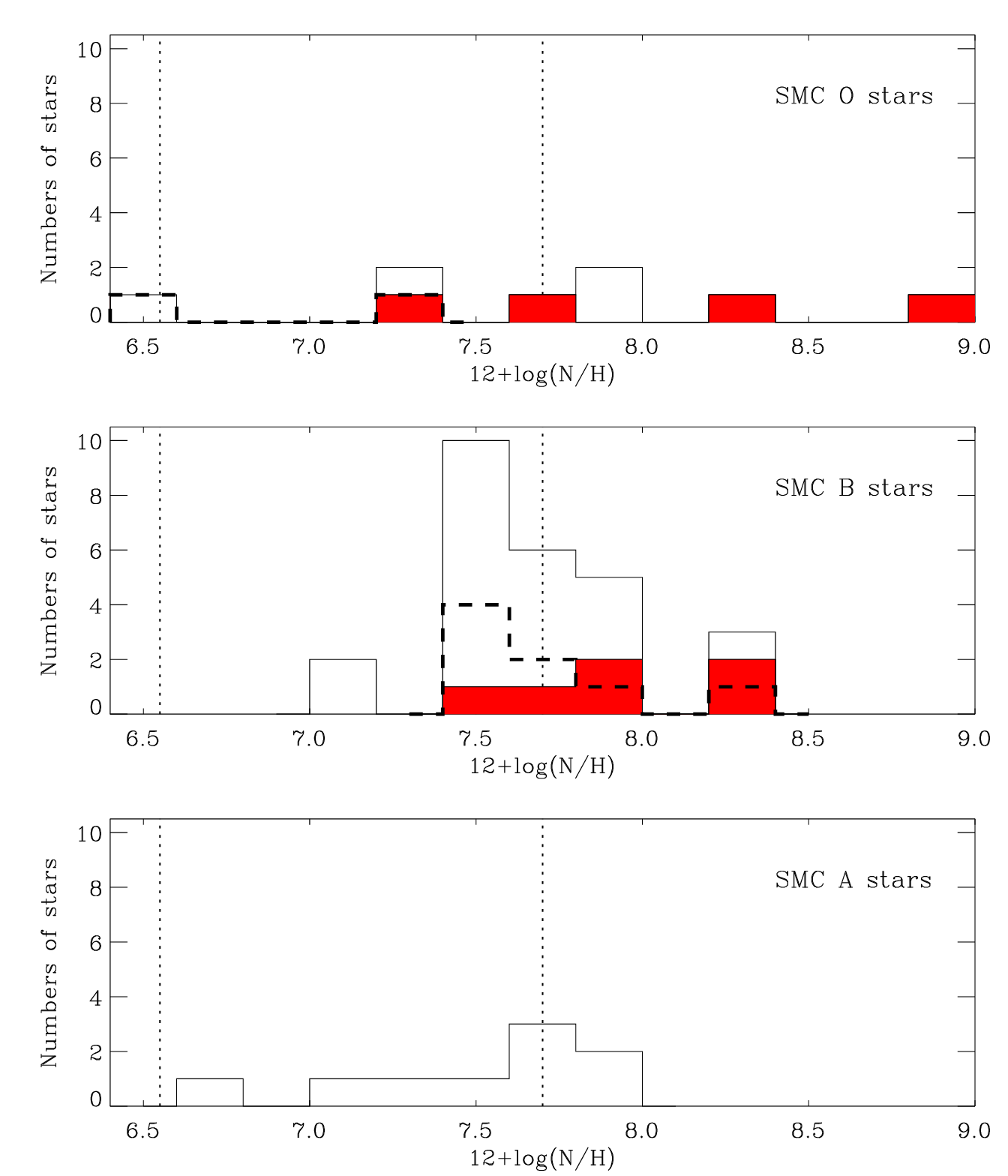}}
}
\caption[]{\label{hist} Histogram of results from abundance analyses of 
SMC OBA-type stars collected from the literature. The dotted lines represent the baseline nitrogen
abundances in the SMC (6.55 dex) and Milky Way (7.70 dex). The top panel shows
the nitrogen abundance of O-type stars in the SMC using the results of
supergiants of Crowther \etal (2002) and Hillier \etal (2003) and SMC O-type
dwarfs of Bouret \etal (2003). The middle panel illustrates the distribution of
observed nitrogen abundances from B-stars in the SMC, these include the results
from this work and the B-type stars of Lennon, Dufton \& Crowley (2003). The 
bottom panel includes the nitrogen abundance results of the A-type
supergiants in the SMC from the analysis of Venn \etal (1999) with the updated
values from Venn \& Przybilla (2003). In the top and middle panel the
distribution of stars with M$_{\rm evol}$ $\sim$ 25 M$_{\odot}$ (box outlined
in bold dashed lines) and 40 M$_{\odot}$ (shaded box) from the original set of
targets are also indicated. }  
\end{figure}

For a comparison of our derived absolute abundances with the results of
previous studies in the SMC, we will assume that the H {\sc ii} region
abundances of  Kurt \etal (1999) represent the present day composition of
the SMC. In addition we include the results of Rolleston \etal (2003;
corrected for non-LTE effects as discussed in Paper 1) for the main-sequence
B-star, AV304, whose photosphere appears to be uncontaminated by any by
products of nuclear processes in the stellar core. Thus, it is assumed that the
chemical make-up of these objects represent the initial baseline composition of
the SMC (for more on this topic see paper 1). The mean element abundance for
this sample is presented in Table~\ref{smc2abund}, but for the purpose of
further discussion the results from all 17 supergiants and 1 giant will be
considered. Hence Table~\ref{smc2comp} presents the mean abundances calculated
by combining the results in Table~\ref{smc2abund} and those presented in Table
5 of paper 1. The results from SMC A-type supergiants and other B-type
stars are also included in Table~\ref{smc2abund} (Venn 1999; Venn \& Przybilla
2003; Lennon, Dufton \& Crowley 2003, these are corrected for non-LTE effects
see Paper 1).

A close examination of Table~\ref{smc2comp} reveals that the $\alpha$-processed
elements (O, Mg, \& Si) are in agreement for all objects, with the possible
exception of the NGC330 stars. The abundances of the $\alpha$-processed
elements in the NGC330 stars appear to be lower than the other SMC objects by
$\leq$ 0.25 dex. This cluster was found to be 0.5 dex metal deficient with
respect to SMC field stars in a photometric study made by Grebel \& Richtler
(1992). However the results here concur with the lower metal deficiency of this
cluster observed in recent analyses by Hill (1999) and Lennon, Dufton \&
Crowley (2004) on K and B-type stars, respectively. CN processed material is
once again evident in the atmospheres of the B-type stars, the nitrogen
abundances in the overall sample studied here varies from 7.14 dex in the
giant, AV216, to 8.30 dex in the most luminous object in the sample, AV78.  The
lowest nitrogen enhancement in the supergiant sample is a factor of 7 (7.40
dex) above the nitrogen abundance of AV304 and the nebular results. This
enrichment is in close agreement with the results from the A-type supergiants
and B-type main-sequence stars (Venn 1999; Venn \& Przybilla 2003; Lennon,
Dufton, \& Crowley 2003). 

To understand the distribution of nitrogen abundances in OBA-type stars in the
SMC, a histogram was assembled from the results in the literature (see
Fig~\ref{hist}). In addition to the results from the present study, the plot
includes the abundances derived for OBA-type stars by Venn (1999), Crowther
\etal (2002), Lennon, Dufton, \& Crowley (2003), Hillier \etal (2003), Venn \&
Przybilla (2003) \& Bouret \etal (2003). The histogram clearly illustrates
that there is a large range in nitrogen abundances observed in these stars. The
majority of the objects have enhanced abundances, which predominantly lie
around $\sim$ 7.50 dex, closer to the baseline galactic nitrogen abundance than
that of the SMC. The stars considered have stellar masses in the range 11 to
60 M$_{\odot}$, thus to decipher any correlation with mass and subsequently
luminosity the results for the 25 M$_{\odot}$ (box outlined in bold dashed line) and 40
M$_{\odot}$ (shaded box) OB stars have been over plotted in Fig~\ref{hist} (no
abundances for A stars in this mass range are available in the literature).
Whilst there appears to be a trend of nitrogen abundance with mass, in that
higher nitrogen abundances are produced in the most massive stars, it is unwise
to make such definitive conclusions from the poor statistics available to us at
present.

Recall that the signature of photospheric contamination by CN-cycled material
is an enhancement in nitrogen with some depletion of carbon and possibly
oxygen. No obvious sign of depletions in oxygen have been detected in this
sample of 18 stars, consistent with theoretical predictions that indicate moderate
depletions in oxygen, smaller than the errors in our analysis. It is
interesting to note that AV215 and AV78 (which have the highest N enhancements 
of 26 and 56 times the baseline value, respectively), have the lowest oxygen
abundances, but these values are still within the error on the mean. In
addition, the mean carbon abundance of the B-type supergiant sample and the
NGC330 stars, are lower than that of AV304 and the H {\sc ii} regions, whether this
is a significant difference is difficult to ascertain given the uncertainties
in the absolute abundances of this ion. Within the B-type supergiant sample
there is no evidence that AV216, (the giant and hence the least processed), has a
higher carbon abundance than the rest of the sample, nor that AV215 and AV78
have depleted carbon. 

\subsection{Evolution of B-type Supergiants}
\label{evol2}

\begin{figure}
\center{
\resizebox{\hsize}{70mm}{\includegraphics{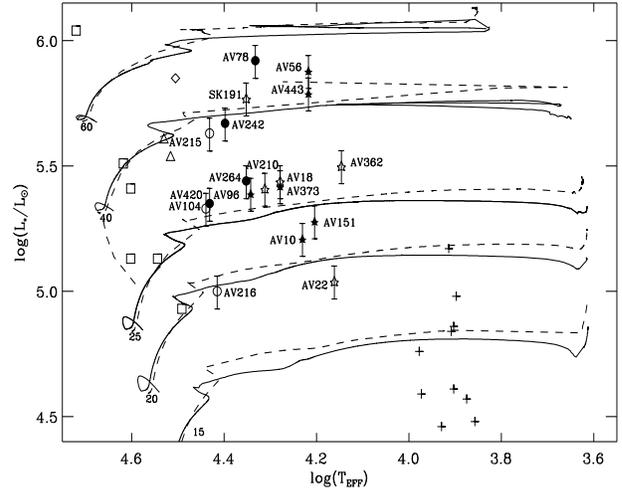}}
}
\caption[]
{\label{hrdiag2} HR-diagram, luminosity as a function of temperature.  The stellar
evolution tracks of Maeder \& Meynet (2001) are shown for an assumed initial
rotational velocity of 0 (---; solid lines) and  300 kms$^{-1}$ (- -; dashed
lines) at solar masses of 15, 20, 25, 40, \& 60 M$_{\odot}$. Included are the
results from the early and mid  B-type supergiants of both the {\sc uves} ($\circ$
\& open star) and {\sc emmi} ($\bullet$ \& $\star$) data, with errors in
luminosity representing $\pm$ 15 \% . In addition, the SMC O-type supergiants of
Hillier \etal ($\triangle$; 2003) and Crowther \etal ($\diamond$ ; 2002), SMC
O-type dwarfs of Bouret \etal (open square ; 2003) and AF-type supergiants from Venn
\etal ($+$; 1999, updated by 2003) are also shown.  
} \end{figure}

In paper 1 it was shown that stellar evolution models which
include the effects of rotation can reproduce, for the most part, the observed
nitrogen abundances in the SMC B-type stars. For the purpose of this
discussion, the complete sample will again be compared to the evolution tracks
of Maeder \& Meynet (2001) which have been re-calibrated to a metallicity of 0.2
Z$_{\odot}$ (see Paper 1) and assume an initial rotational velocity of
300 kms$^{-1}$. The position of this stellar sample on the HR diagram is illustrated
in Fig~\ref{hrdiag2}. From these tracks all of the stars seem to be past the
end of the hydrogen burning main-sequence phase, with some of the objects
having thick hydrogen burning shells. Indeed the loci of the cooler stars
(B2.5 - B5) on the evolutionary tracks imply that they are in the core
helium-burning phase.  

In these comparisons one must be careful as the actual stars may have a range
of initial rotational velocities rather than the 300 \km assumed in the
evolutionary models, and therefore the tracks presented in Fig~\ref{hrdiag2}
may be inappropriate for some, or all, of the stars. Including rotational
velocities in the evolutionary models has the effect of extending the
main-sequence branch to higher luminosities and to lower temperatures, which
will therefore result in lower evolutionary masses for a star of a given
luminosity. Stars with initial stellar masses $>$ 15 M$_{\odot}$ tend to cross
the HR-diagram from the end of the main-sequence phase towards cooler
temperatures with relatively constant luminosities, on the way to becoming red
supergiants (RSG). Maeder \& Meynet (2001) showed a comparison of evolutionary
tracks adopting initial rotational velocities of 0, 200, 300 and 400 \km, which
implies only a small difference in luminosity ($\Delta\log
(L_{\odot}/L_{\star})  < $ 0.1) as the star moves across the HR-diagram for an
initial stellar mass of M$_{\star}$ = 20 M$_{\odot}$ (see Fig. 8 in Maeder \&
Meynet 2001). It is therefore reasonable to estimate evolutionary masses from
these tracks, with the caveat that the assumed initial rotational velocities
and the errors in interpolating between the tracks lead to uncertainties of 
$\sim$ 5 solar masses (and possibly more at the high end of the mass range
under consideration).

\begin{figure}
\center{
\resizebox{\hsize}{70mm}{\includegraphics{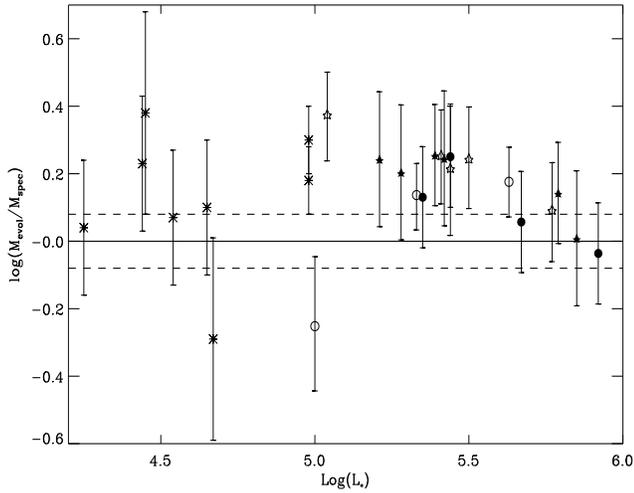}}
}
\caption[]{\label{mass2}
Comparison of evolutionary and spectroscopic masses as a function of
luminosity.  Included are the results from the early and mid B-type supergiants
of both the {\sc uves} ($\circ$ \& open star) and {\sc emmi} ($\bullet$ \&
$\star$) data. Additionally the results from NGC330 B-type stars from Lennon,
Dufton \& Crowley ($\ast$;  2003) are shown. The error bars represent the error in
the surface gravity. Also shown is the uncertainty in mass due to the adopted
distance modulus (- -; dashed lines).  
} \end{figure}

The estimated evolutionary masses, M$_{\rm evol}$, based on the 300 \km tracks
are displayed in Table~\ref{smc2par3} along with the spectroscopic masses,
M$_{\rm spec}$. These exhibit the same behaviour as found from the analysis of
the {\sc uves} data, such that M$_{\rm evol}$ $>$ M$_{\rm spec}$ by
approximately 0.10 - 0.30 dex (see Fig.~\ref{mass2}). The errors in the adopted
distance and gravity account for much of the discrepancy, leaving only a
difference of 0.10 dex (or $\sim$ 8 M$_{\odot}$). It is difficult to account for
the remaining disagreement unless it is an overestimation of M$_{\rm evol}$ due
to errors in interpolation. However, the largest discrepancies are observed in
the less luminous, and hence less massive stars. In these stars the tracks are
more widely spaced and therefore more accurate estimates of M$_{\rm evol}$
would be expected. By including some degree of convective overshooting the
track for a given mass would be extended to higher luminosities and could
account for at least some, if not all, of the overestimation in M$_{\rm evol}$.
Another possible explanation is that the lower luminosity stars may be in a
post-RSG blue-loop phase and that the lifetime for this phase is greater than
the pre-RSG stage.

\begin{figure}
\center{
\resizebox{\hsize}{70mm}{\includegraphics{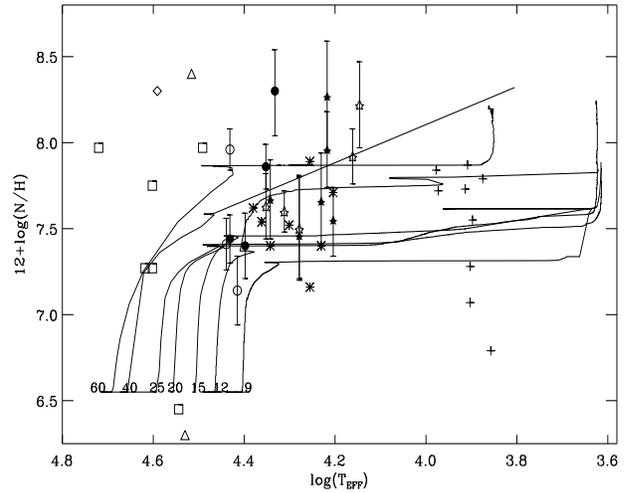}}
}
\caption[]
{\label{Nteff2} Surface nitrogen abundances as a function of temperature. The
stellar evolution tracks of Maeder \& Meynet (2001) are shown for an assumed
initial rotational velocity of 300 kms$^{-1}$ and various solar masses.
Included are the results from the early and mid B-type supergiants of both the
{\sc uves} ($\circ$ \& open star) and {\sc emmi} ($\bullet$ \& $\star$)
data. Also displayed are the results from SMC O-type supergiants of Hillier
\etal ($\triangle$ ; 2003) and Crowther \etal ($\diamond$ ; 2002), SMC O-type
dwarfs of Bouret \etal (open square; 2003), NGC330 B-type stars of Lennon, Dufton,
\& Crowley ($\ast$ ; 2003), and A-type supergiants from Venn \etal (+; 1999,
updated by 2003). The error bars illustrate the random and systematic errors
on the nitrogen abundance.  
} 
\end{figure}

To appreciate the success of the new evolutionary models, which consider
rotation, in reproducing the observed nitrogen abundances in hot, luminous
stars it is useful to look at the production of nitrogen as a function of
temperature for various stellar masses (see Fig.~\ref{Nteff2}). The models with
stellar masses in the range  20 - 60 M$_{\odot}$ have nitrogen enrichments by
the end of the core hydrogen burning phase of a factor of 7 to 20 above the
baseline nitrogen abundance of 6.55 dex (from AV304). This is similar to the
enhancements observed in our sample, although some of the more luminous objects
have greater enhancements. Indeed, adopting a stellar mass  for each object from
the HR-diagram in Fig~\ref{hrdiag2} and following the appropriate evolution
track in Fig.~\ref{Nteff2}, it would appear that many of the objects have
greater enhancements than predicted. However given the errors in the abundance
determinations this may not be significant. The two B2 Ia stars (AV373 \& AV18)
are interesting objects as they have very similar atmospheric and wind
parameters, in addition their nitrogen abundances are the same, \emph {within
the errors}. This is also the case for the two B0.5 Ia stars, AV420 and AV104.
Indeed the position of these four stars on the HR-diagram fall close to the
theoretical predictions for a 25 M$_{\odot}$ star and have abundances in
agreement with the models for a 25 M$_{\odot}$ star with initial rotational
velocities of 300 \km.

From line driven wind theory, it is predicted that the stellar winds are
stronger in stars of higher metallicity (\ie that the mass-loss rates are
higher). This comes directly from the dependence of the photon momentum
transfer to the wind on the number of optically thick lines and  that at
lower metallicities the line strengths decrease. If this is the case, the loss
of angular momentum by the end of the main-sequence is much greater in galactic
objects than in lower metallicity stars (viz. SMC). This reduction in angular
momentum at for example, solar metallicity, leads to a more rapid decrease in
stellar rotational velocity. Therefore, the models by Maeder \& Meynet (2001)
for stars with Z = 0.004 (SMC metallicity) have higher rotational velocities
after core hydrogen burning than similar stars with solar metallicity. Since
these stars have higher rotational velocities the convective zones are larger
and the atmospheric distortions more severe, this results in greater
enhancements in the nitrogen content of the photosphere. 

\emph {Do we observe this variation in rotational velocities from evolved stars
in the two different environments?} Unfortunately this question is difficult to
answer based on spectra alone. We do see greater nitrogen enrichments in the SMC
stars, yet it is not possible to ascribe this to rotational velocities at
present. $v\sin i$ values have been measured for OBA-type stars in the galaxy
and SMC but the estimates for both regions are similar (in the range 40-150
\km). However as noted previously in Sect.~\ref{obs2}, these measurements are
not truly representative of the projected rotational velocity but are dominated
by macroturbulence. Also note that these are the projected rotational
velocities and some information on the inclination of the rotational axis is
required to enable the true velocity to be disentangled from $v\sin i$
measurements.

\subsection{Mass-loss: Observations and Predictions.}
\label{wind2}

In this section the results of the entire sample of B-type supergiants will be
discussed in an attempt to glean more information on the behaviour of their
stellar winds as a function of luminosity, spectral type and metallicity. From
a similar analysis of B-type stars in the Galaxy, Kudritzki \etal (1999) found
a clear distinction between the wind momenta in early and mid B-type stars.
This behaviour also differed from that of the O-type stars from Puls \etal
(1996).  This change in wind-momenta with spectral type is dependent on the
flux weighted-distribution of the spectral lines, the ionisation of the lines
contributing to the line force and the relative number of strong to weak lines
driving the wind. The change in wind-momenta observed in B-type stars by
Kudritzki \etal occurred between two objects with spectral types B1 and B1.5 at
\teff = 23.5 and 22.5 kK (note these temperatures were determined from
unblanketed models). This is approximately the point in which the so-called
'bistability jump', a change in the properties of the wind, is predicted to
take place (Lamers, Snow \& Lindholm 1995). 

Due to the range in temperatures observed for the B1 stars in this sample and
the fact that it overlaps with that of the B1.5 stars, it is more appropriate
to make the early/mid B distinction based on stellar temperatures rather than spectral
type. The hottest B1.5 star in this sample (Sk191) has an effective temperature
of 22.5 kK. In Paper 1 it was shown that this object has a wind which behaves
similar to the cooler stars, therefore this has been adopted as the cut off
temperature for early B-type stars in the following discussion. This implies
that AV264 and AV78 may exhibit signs of having lower wind momenta than the
hotter B1 stars, but similar to that of the mid B-type stars. A direct
comparison of the wind momenta for these stars with those of the hotter B1-type
objects is not appropriate due to the luminosity dependence of the stellar
wind. 

\begin{figure} 
\center{
\resizebox{\hsize}{70mm}{\includegraphics{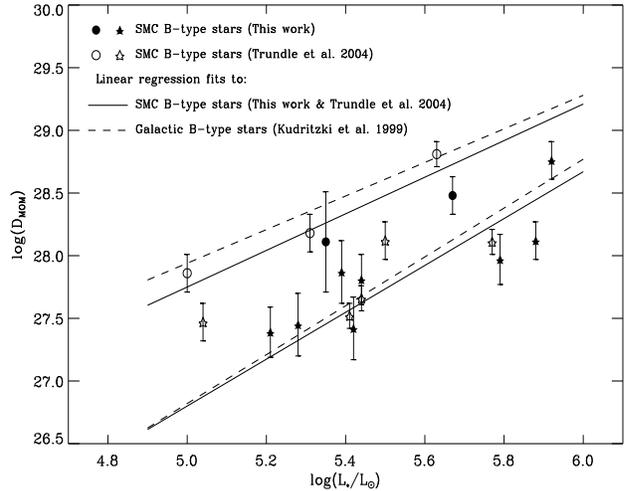}}
}
\caption[]{\label{wlr3} WLR as derived from SMC B-type supergiants.  The plot
shows the early (circles) and mid (stars) B-type stars from the  SMC
sample({\sc uves}: open symbols; {\sc emmi}: filled symbols). Included are the linear
regression fits to the observational results of the early (all) and mid (only $<$
B3) B-type stars (----; solid lines). The dashed lines (- - -) are the
wind-momentum-luminosity relationship, from galactic B-type supergiants without
line-blanketing (Kudritzki \etal 1999). The error bars represent the
uncertainty in deriving the wind momentum.}  
\end{figure}

No clear distinction between the wind-momenta of the SMC sample and the WLR of
their galactic counterparts is evident. This consistency still holds if one
regards the wind-momenta of the individual Galactic and SMC B-type supergiants
at a certain luminosity, although a larger scatter is observed in the SMC mid
B-type stars. It seems appropriate to attempt an estimation of the WLR from
these SMC targets since they are the first homogeneous results from B-types
stars in a low metallicity environment.  We adopt the same form of the 
WLR as defined by Kudritzki \etal (1999)

\begin{equation}
\log D_{\rm mom} = \log D_o + x\log(L_{\star}/L_{\odot})
\end{equation}

\noindent where the inverse of the slope, 1/x = $\alpha^{\prime}$.
$\alpha^{\prime}$  describes the depth dependence of the radiative line force
and can be expressed as a function of the force multiplier parameters of the
radiative line force (i.e $\alpha^{\prime} = \alpha - \delta$). D$_{\rm mom}$
is a product of the mass-loss rate, terminal velocity of the wind and
quadrature of the stellar radius. 

The linear least squares fit to these results, which take into
consideration the errors of both variables, can be seen in Fig~\ref{wlr3} and
the coefficients ($\log$ (D$_{\rm o}$) and x) are given in
Table~\ref{smc2wlr}.  The galactic WLR from Kudritzki \etal (1999) is also
shown, however these did not include any uncertainty in luminosity which may
taint the comparison with the SMC WLR. For the early B-type supergiants, the
wind-momentum at a given luminosity is between 0.08 to 0.18 dex lower for the
SMC stars than the Galactic stars. This difference is dependent on the
luminosity due to the steeper slope of the SMC relationship.  Theoretical and
observational evidence suggests that stars with different luminosity
classifications have wind momenta which vary (Puls \etal 1996; Puls, Springmann
\& Lennon 2000). Therefore it may be more appropriate to consider the WLR for
only the four supergiants, omitting the giant AV216. The regression
coefficients presented in Table~\ref{smc2wlr} reveal that this results in a
steeper WLR with $\alpha^{\prime}$ = 0.52, which is lower than that obtained
from the Galactic B stars. However one must be cautious in interpreting these
results as it is still a statistically poor sample. 

\begin{table*}
\caption[Line coefficients for WLR]
{\label{smc2wlr} The coefficients for the linear regression
fits to the observed and theoretical predictions (Vink et al. 2001; V01)
 of the
WLR from the SMC supergiants in this sample (all). Coefficents
describing the relationship determined from the 4 early B-type
supergiants, omitting the giant AV216, are also given 
(sg). In
addition the values describing the relationship of only the
mid B-type stars with T$_{\rm eff}$ in the range 22.5 - 16 kK
are given (only $<$ B3). Included are the coefficients for the
Galactic B-type supergiants analysed by Kudritzki \etal (1999;
K99) and that determined from the early B-type supergiants
analysed by Urbaneja (2004; U04). }
\begin{flushleft}
\centering
\begin{tabular}{lccc} \hline \hline
Analysis  &  $\log$ (D$_{\rm o}$) & x & $\alpha^{\prime}$ 
\\
\hline
          &\multicolumn{3}{c}{Early B-type stars}
\\ 
\hline
SMC This work (all) & 20.45 $\pm$ 2.54 & 1.46 $\pm$ 0.44 & 
0.68 $\pm$ 0.15 
\\
SMC This work (sg)   & 17.91 $\pm$ 6.11 & 1.91 $\pm$ 0.93 & 
0.52 $\pm$ 0.43 
\\
Galactic K99           & 21.24 $\pm$ 1.38 & 1.34 $\pm$ 0.25 & 0.75 $\pm$ 0.15 
\\
Galactic U04           & 21.72 & 1.31  & 0.76
\\
SMC V01                & 16.67 & 2.03  & 0.49 
\\
\hline
                       &\multicolumn{3}{c}{Mid B-type stars}
\\
\hline 
SMC This work (all) & 18.39 $\pm$ 2.57 & 1.72 $\pm$ 0.44 & 
0.58 $\pm$ 0.16 
\\
SMC This work (only $<$ B3) & 17.45 $\pm$ 3.14 & 1.87 $\pm$ 
0.53 & 0.53 $\pm$ 0.16
\\
Galactic K99    & 17.07 $\pm$ 1.05 & 1.95 $\pm$ 0.20 & 0.51 $\pm$ 0.05
\\
SMC V01         & 18.33 & 1.87 & 0.53
\\
\hline 
\\
\end{tabular}
\end{flushleft}
\end{table*} 

The coefficients for the mid B-type stars are also included in
Table~\ref{smc2wlr}. Considering all stars with \teff $<$ 22.5kK, results in a
slightly shallower WLR than obtained for the Galactic stars. In paper 1
it was suggested that the cooler stars of the sample AV362 and AV22 (B3 and
B5-type stars respectively) may exhibit a different behaviour to that of the
mid B-types, possibly similar to that of the A-types. If this is the case then
the mid B-type WLR should be determined only from those stars with 22.5 $>$
\teff $>$16kK. A steeper relationship is then obtained with 
$\alpha^{\prime}$ = 0.53, which is in very good agreement with the
Galactic stars. 

As mentioned earlier, Puls, Springmann \& Lennon (2000) concluded from a
theoretical study of the line forces driving the winds in massive stars that in
lower density winds, lower $\alpha$ and hence $\alpha^{\prime}$ exist. An
interesting point is that this also suggests that a lower exponent is expected
for low metallicity environments where the winds should be less dense, \ie that
$\alpha^{\prime}$ would be smaller at the metallicity of the SMC compared to
stars with solar abundance. This is the case for the early-type stars but not
for the mid B objects, yet the situation for the latter may be unclear as
many of the Galactic objects are classed as Ib whilst those in the SMC sample
are all more extreme type-Ia stars.

\begin{figure} 
\center{
\resizebox{\hsize}{70mm}{\includegraphics{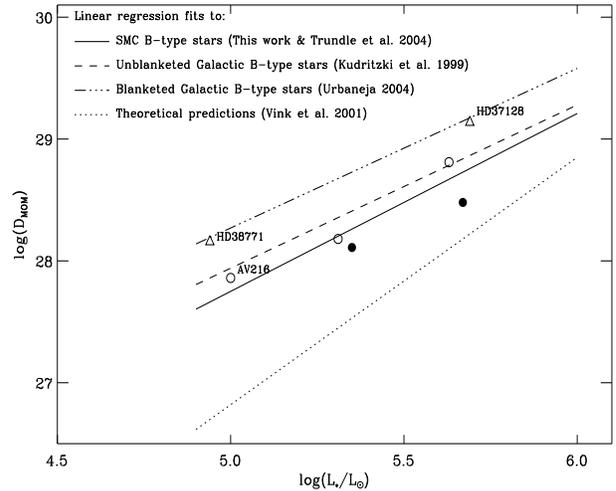}}
}
\caption[]{\label{wlr5} Calibration of WLR with metallicity  for SMC early
B-type supergiants.  The plot shows linear regression fits to the early B-type
stars from the SMC sample (solid line with {\sc uves}: $\circ$ ; {\sc emmi}: $\bullet$) and
Galactic stars analysed with a blanketed version of {\sc fastwind} (Urbaneja
2004; dash-dotted line with $\triangle$). Also included are the WLR's from
theoretical predictions based on Vink \etal (2001; dotted line) and unblanketed galactic
results from Kudritzki \etal (1999; dashed line).} 
\end{figure}

\begin{figure}
\center{
\resizebox{\hsize}{70mm}{\includegraphics{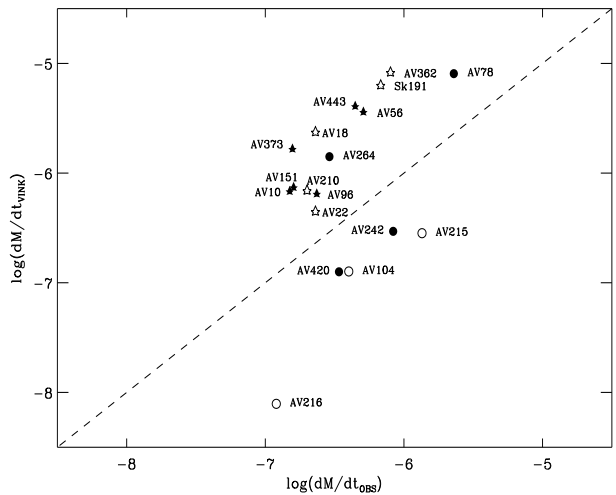}}
}
\caption[]
{\label{vink2} Comparison of theoretical and observed mass-loss rates.
Theoretical predictions are calculated using the metallicity dependent
mass-loss recipes of Vink \etal (2001) for Z = 0.2 Z$_{\odot}$. The
circles ({\sc uves}: $\circ$; {\sc emmi}: $\bullet$) represent the SMC early
B-type supergiants,  whilst the stars ({\sc uves}: open star; {\sc
emmi}: $\star$) represent the mid B-type supergiants. }
\end{figure}

\begin{figure}
\begin{center}{
\resizebox{\hsize}{70mm}{\includegraphics{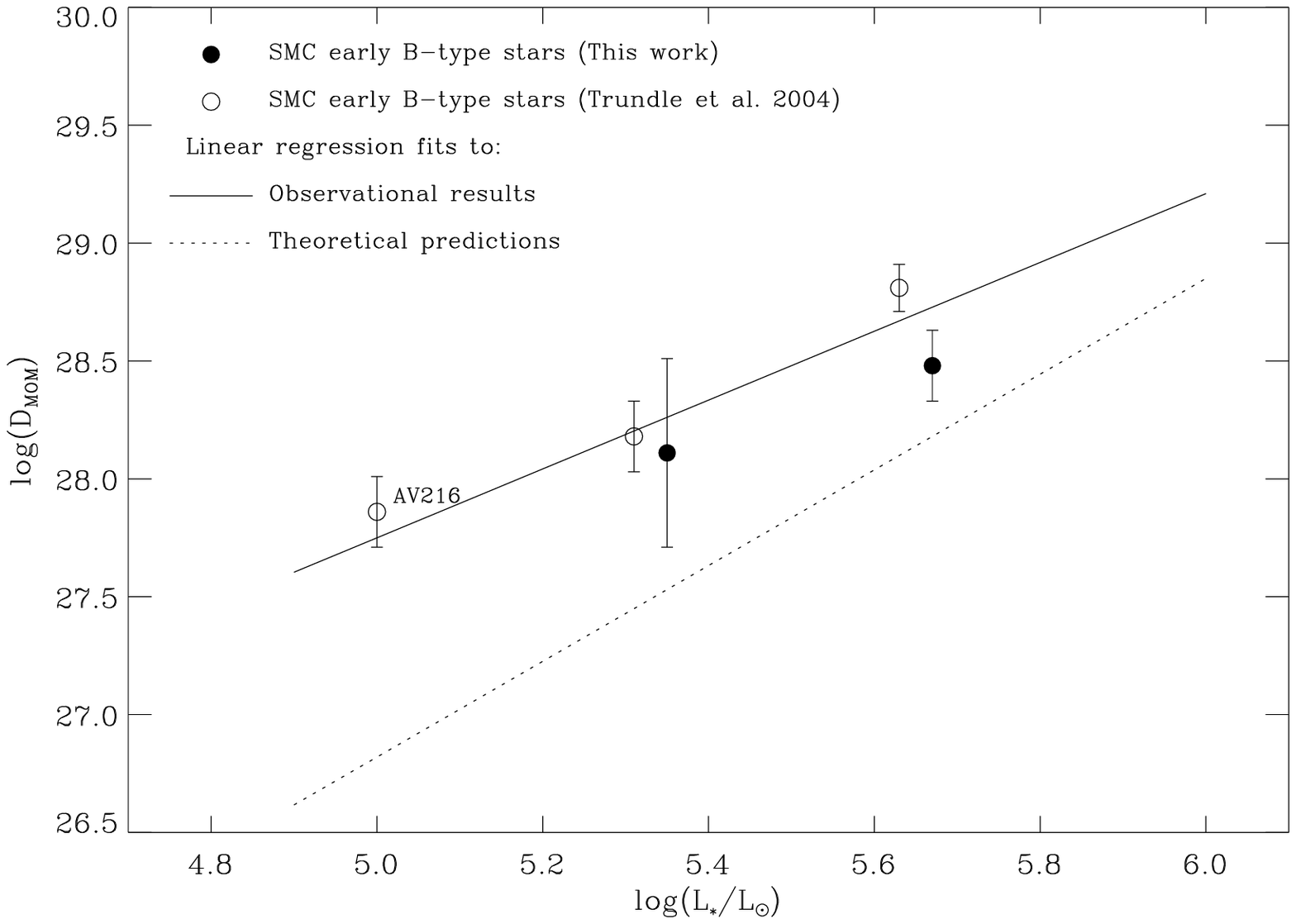}}
\resizebox{\hsize}{70mm}{\includegraphics{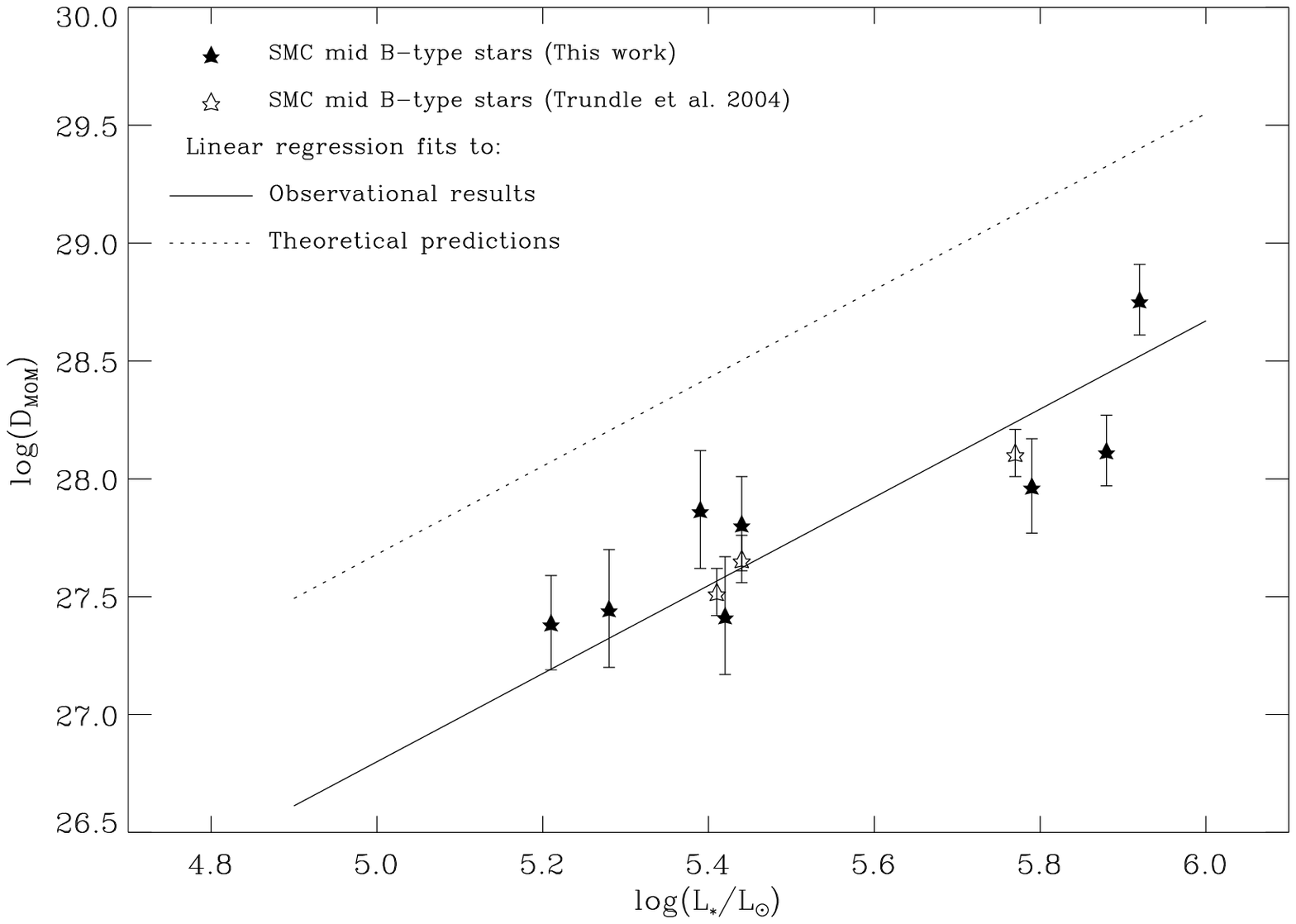}}
}
\caption[Comparison of theoretical and observational wind momenta]
{\label{wlr4} Comparison of theoretical and observational wind momenta for
B-type supergiants in the SMC. In the upper panel the observational result for
the early B-type supergiants ({\sc uves}: $\circ$; {\sc emmi}: $\bullet$) from
this work are considered. Included are the linear regression fits to the
observational results (-----; all) and the predicted wind-momenta of Vink \etal
(2001; $\cdot\cdot\cdot\cdot\cdot$). In the lower panel the observational
result for the mid B-type supergiants ({\sc uves}: open star; {\sc emmi}:
$\star$) from this work are plotted together with the linear regression fits
from the observational (only $<$ B3) and theoretical results. The error bars 
represent the uncertainty in deriving the wind momentum.}
\end{center}
\end{figure}

An additional effect to be considered in our interpretation of the results, is
that the analysis carried out by Kudritzki \etal (1999) was based on
unblanketed {\sc fastwind} models. This results in lower wind-momenta than
derived from blanketed models (see effect in Repolust, Puls, \& Herrero 2004). 
Hence the true variation in wind-momenta will probably be found to be larger
when compared to results from similar model atmosphere analyses.  This may
explain why no significant metallicity dependence is evident here. In order to
investigate this, a preliminary Galactic WLR has been determined from two
Galactic early B-type objects analysed by Urbaneja (2004) using a line
blanketed version of {\sc fastwind}. Indeed from  Fig~\ref{wlr5} a significant
difference of $\sim$ 0.5 dex is observed between this relationship and that
derived from the SMC objects. Although this is only derived from two
objects, it supports the theoretical prediction for a dependence of radiative
driven winds on metallicity.

In Paper 1 the mass-loss rates derived from the observational data were
discussed in relation to theoretical predictions from Monte-Carlo simulations
(Vink \etal 2000, 2001). From the analysis of the {\sc uves} data, a
discrepancy was found between the observed mass-loss rates and theory; the
theoretical predictions were lower for the early B and higher for the mid
B-type stars by a factor of 5 and 7, respectively. A comparison of the entire
sample of observed mass-loss rates with the theoretical predictions of Vink
\etal (2001) for a metallicity of Z=0.004 (0.2 Z$_{\odot}$) is illustrated in
Fig.~\ref{vink2}. It is immediately evident from this plot that the behaviour
of the {\sc emmi} sample is similar to that of the {\sc uves} dataset; the
theoretical mass-loss rates of the early-type stars are under estimated and
those of the cooler stars are over estimated in relation to the observed values
(Note that the momentum driving the winds of AV78 and AV264 appear to follow
the same pattern as the mid B-type stars). The Vink \etal (2000) formula also
over estimates mass-loss rates for the Galactic mid B-type supergiants analysed
by Kudritzki \etal (1999; this is illustrated in Fig 10 of Vink \etal 2000). It
was shown by Vink \etal that \.M derived from radio observations by Scuderi
\etal (1998) appeared to be in agreement with the predictions for mid B-type
supergiants. These two studies, also agreed with \.M estimates from
H$_{\alpha}$ lines in pure emission but not with those derived with P-Cygni or
absorption profiles. In the complete SMC sample considered here, seven out of
thirteen mid B-type stars display pure emission and they do not appear to have
systematically higher mass-loss rates than those derived from P-Cygni or
absorption profiles. Therefore, the discrepancy between the Vink predictions
and the observational results presented here does not appear to be a
consequence of the H$_{\alpha}$ profile. 

This discrepancy with the theoretical predictions naturally results in a 
similar discrepancy in the wind-momenta. Adopting the predicted mass-loss rates
and observed terminal velocities of the stars, the wind-momenta were
determined. Fig~\ref{wlr4} illustrates the significant difference between the
linear regression fits to the observed and theoretical wind-momenta. The
coefficients for the theoretical predictions are included in
Table~\ref{smc2wlr}. An important point to note is that while the observed
wind-momenta for the early-type objects are higher than the mid B-type objects
the opposite behaviour is predicted from the theoretical relationships  
(\ie the mid B-type stars are predicted to have greater wind-momenta than their
hotter counterparts). This is a substantial disagreement between the
observational and theoretical results which may arise from the assumptions in
either the model atmosphere analysis or the theoretical study of Vink \etal
(2001). As discussed in Paper 1, the consideration of clumping in the wind
can reduce the derived mass-loss rates and this would move the observational
results of the early B-type objects into better agreement with the theoretical
predictions. In contrast, clumping will increase the discrepancy with the mid
B-type stars. Whilst a simple form of clumping can now be included in the
unified model atmosphere codes such as {\sc cmfgen} and {\sc fastwind},
invoking such processes is still largely speculative without  knowledge
of the clumping in the wind along our line of sight. In addition one must be
aware of the variability in the winds of massive stars which time series optical
(e.g. H$_{\alpha}$ profile) and UV (e.g. P-Cygni lines) observations of galactic
objects have revealed in recent years (for example Prinja, Massa \& Fullerton 2002; Kaufer, Prinja \& Stahl 2002; Morel
\etal 2003; Prinja \etal 2004). Whilst the variability in the H$_{\alpha}$
profiles of some of these Galactic objects may correspond to considerable
changes in the properties of the winds, without further observations of the SMC
objects we cannot investigate the extent of their variability (if any) and
the subsequent affect on the derived wind parameters.

The theoretical predictions for the SMC objects give 0.49 and 0.53 as the value
of the exponent $\alpha^{\prime}$ for early and mid B-type stars, respectively.
This steeper relationship, in comparison to observations, is evident from
Fig~\ref{wlr4}.  The observed value of $\alpha^{\prime}$ has an important
impact on stellar evolution calculations where it is used to describe mass-loss
rates as a function of rotation. The smaller this factor the greater the
mass-loss rate becomes at a given rotational velocity and position on the HR
diagram. Thus, $\alpha^{\prime}$ has an influence on both the stellar lifetimes
and element yields. In the Maeder \& Meynet (2001) code this exponent is
assumed to be 0.66, which with in the errors agrees well with those found from
the current sample of stars. Nevertheless the large discrepancy between
the mass-loss rates of the mid B-type stars is of concern. Since the large
values predicted by Vink \etal (2001) would produce a rapid spin down as the
star evolves beyond the main-sequence, which would not be predicted by
evolution models if the observed results were invoked in the code. Furthermore
$\alpha^{\prime}$ is important for the accuracy of the distances obtained from
the WLR, since the distance derived is dependent on a precise determination of
the luminosity. The steeper the WLR and hence smaller $\alpha^{\prime}$ the
more accurately the luminosity of an object can be determined from its derived
wind-momenta and subsequently the more useful the WLR technique is as a
distance estimator.

\section{Conclusions}

In summary, a complete unified model atmosphere analysis has been completed for
a further ten B-type supergiants in the SMC. By incorporating the eight stars
analysed in paper 1, this sample of 18 objects has been considered in terms of
the properties of their stellar winds and massive star evolution. 

The nitrogen abundances in this dataset provide ample evidence of enhancement
in the stellar photosphere.  Incorporating the nitrogen abundances available in
the literature for O-type stars and BA-type supergiants, it was shown that the
majority of these have significantly enhanced nitrogen, with an average of 7.5
dex (\ie close to solar abundance). This degree of  nitrogen enrichment can be
reproduced by stellar evolution models with initial rotational velocities of
300 \km (Maeder \& Meynet 2001). However, the high rotational velocities these
models predict at the end of the main-sequence do not appear to be consistent
with the observed projected rotational velocities.

A calibration of the wind-momenta as a function of luminosity for the B-type
supergiants was considered, taking a step closer to the use of the WLR
relationship as a distance indicator. The SMC relationships are inconsistent
with the Vink \etal (2001) predictions, which for both the early and mid B-type
objects are steeper. A comparison with two galactic early B-type stars analysed
by Urbaneja (2004), revealed clear evidence for the metallicity dependence of
radiatively driven winds.

Despite this, the spread in luminosity and temperature ranges possible for a
given spectral type requires more extensive samples to disentangle the
behaviour of momenta with these properties. In the future it is attractive to
use larger samples to tie down this behaviour and determine $\alpha^{\prime}$;
an important factor for stellar evolution codes.


\section*{Acknowledgements}   

CT is grateful to the Department of Higher and Further Education, Training and
Employment for Northern Ireland (DEFHTE) and the Dunville Scholarships fund for
their financial support. DJL acknowledges funding from the UK Particle Physics
and Astronomy Research Council ({\sc pparc} under the grant PPA/G/S/2001/00131.
We would like to thank Joachim Puls and Robert Ryans for their continuous
support with {\sc fastwind} and {\sc tlusty}, respectively. We are also
grateful to Miguel Urbaneja for providing the galactic supergiant results 
and the referee, Alex Fullerton, for his useful comments.%

%

\end{document}